\documentstyle[preprint,eqsecnum,aps,axodraw]{revtex}
\tighten
\begin{document}
\title{Perturbative S-matrix in discretized light cone quantization of 
two dimensional $\phi^4$ theory}
\author{{\bf A. Harindranath}$^{a,b}$, {\bf L$\!\!$'. Martinovi\v c}$^{a,c}$, and 
{\bf J. P. Vary}$^{a}$ \\
$^{a}$ Department of Physics and Astronomy, Iowa State University, Ames, IA
5001, U.S.A. \\
$^{b}$Saha Institute of Nuclear Physics, 1/AF Bidhan Nagar, Calcutta, 700064,
India \\
$^{c}$ Institute of Physics, Slovak Academy of Sciences \\
D\'ubravsk\'a cesta 9, 842 28 Bratislava, Slovakia\\}
\date{ January 29, 2002}
\maketitle
\begin{abstract}
We study the S-matrix of two-dimensional $\lambda\phi^4$ theory in 
Discretized Light Cone Quantization and show how the correct 
continuum limit is reached for various processes in lowest order perturbation
theory.  
\end{abstract}
\vspace{0.5cm}
\vspace{0.5cm}
%%%%%%%%%%%%%%%%%%%%%%%%%%%%%%%%%%%%%%%%%%%%%%%%%%%%%%%%
\section{Introduction}
%%%%%%%%%%%%%%%%%%%%%%%%%%%%%%%%%%%%%%%%%%%%%%%%%%%%%%% 
S-matrix elements have been studied in the continuum formulation of 
light front quantization since its inception\cite{Bjorken:1971ah}. Chang
and Yan\cite{Chang:1973qi} showed the formal equivalence of S matrix elements in light
front quantization and the more familiar instant form of quantization. Some
doubts were nevertheless raised\cite{Suzuki:1976xb} regarding the formulation of a consistent
scattering theory in the light front formulation. Detailed calculations 
of phase shifts were 
carried out\cite{Ji:1992xr} in $\phi^3$ theory with emphasis on issues
concerning rotational invariance (also see Ref. \cite{Fuda:1991nn}.

Discretized Light Cone Quantization (DLCQ)
\cite{Maskawa:1976ky,Casher:1976ae,Thorn:1978kx} is a method proposed for the
non-perturbative solution of quantum field theories\cite{Brodsky:1998de}. 
Most of the applications of the method following the work of Refs.
\cite{Pauli:1985pv,Pauli:1985ps} have been to bound state
spectra. Only very recently have studies focused on the application of
DLCQ to the calculation of scattering observables\cite{Hiller:2000vi}.

A calculation of one loop scattering amplitude in two dimensional
$\phi^4$ theory was carried out in 
Ref. \cite{Chakrabarti:2000cg} for the $s$-channel process
below production threshold to illustrate how the correct 
continuum limit was approached
for this process in DLCQ. In a recent
work\cite{Harindranath:2000vf} problems associated with
compactification near and
on the light front have been investigated in detail in the context of
perturbative scalar field theory. This work was motivated by the result of
Ref. \cite{Hellerman:1999yu} that certain divergences
arise in the one
loop scattering amplitude in scalar field theory at {\em finite box length}
as one tried to approach the light front in a formalism of compactification
near the light front. By means of detailed calculations in both continuum
and discrete versions in three different
approaches: (1) quantization on a space-like surface close to a light   
front; (2) infinite momentum frame calculations; and (3) quantization on
the  light front, Ref. \cite{Harindranath:2000vf} concluded
that in DLCQ, contributions from 
$``$zero mode (ZM) induced" interaction terms decouple in the continuum limit
and covariant results are reproduced.

However, the claim of Ref. \cite{Harindranath:2000vf} regarding the continuum 
limit of DLCQ for processes with $p^+=0$
exchange has been challenged in a very recent work\cite{Taniguchi:2001cb}.  
Authors of Ref. \cite{Taniguchi:2001cb} agree with the conclusion of 
Ref. \cite{Harindranath:2000vf} that contributions from 
ZM induced interaction terms in DLCQ decouple in the continuum 
limit but they claim that DLCQ yields vanishing forward scattering amplitude 
in the continuum limit whereas the correct result is finite. In view of the
persistent confusion on the subject, it is worthwhile 
to provide details of our simple, straightforward and unambiguous 
calculation and
reconfirm our original claim. 
We show how the careful treatment of the process of taking the continuum
limit in DLCQ yields the correct result. We also provide detailed numerical
results.

The plan of this work is as follows. In Sec. II we present light front
perturbation theory calculation of one loop scattering amplitude in ${
\lambda \over 4!} \phi^4$ theory in the continuum formulation and
discretized formulation and numerical results. Sec. III contains discussion,
summary, and conclusions.  
%%%%%%%%%%%%%%%%%%%%%%%%%%%%%%%%%%%%%%%%%%%%%%%%%%%%%%%%%%%%
\section{Light front perturbation theory calculation of 
one loop scattering in ${\lambda \over 4!} \phi^4$ theory
}
%%%%%%%%%%%%%%%%%%%%%%%%%%%%%%%%%%%%%%%%%%%%%%%%%%%%%%%%%%%%%
\subsection{Continuum formulation}
%%%%%%%%%%%%%%%%%%%%%%%%%%%%%%%%%%%%%%%%%%%%%%%%%%%%%%%%%%%%%%%%%%%%%
%%%%%%%%%%%%%%%%%%%%%%%%%%%%%%%%%%%%%%%%%%%%%%%%%%%%%%%%%%%%%%
\subsubsection{$t$-channel scattering}
%%%%%%%%%%%%%%%%%%%%%%%%%%%%%%%%%%%%%%%%%%%%%%%%%%%%%%%%%%%%%%
Let us first review the forward scattering limit in the continuum
formulation. For simplicity we will consider two dimensional theory since
extra dimensions do not add or subtract to the essential features and
conclusions of the calculation. 

Consider the scattering amplitude at one loop level in $ \phi^4$ theory.
$p_1,p_2$ are the initial momenta and $p_3,p_4$ are the final momenta. 
 Let us denote $s= (p_1+p_2)^2$ and $ t = (p_1-p_3)^2$. 
In the light front perturbation theory, we have to consider two 
cases separately. 

1) $p_1^+ > p_3^+$. 

The scattering amplitude (Fig. 1a) is
\begin{eqnarray}
M_{fi} && = {1 \over 2}{ \lambda^2 \over 4 \pi} ~\theta(p_1^+ - p_3^+)~
\int_0^{p_1^+ - p_3^+} 
dq_1^+ 
 ~{ 1 \over q_1^+} ~{ 1 \over p_1^+ - p_3^+ - q_1^+}
\nonumber \\
&&~~~~~~~~~{ 1 \over p_1^- + p_2^- - p_3^- -p_2^- - q_1^- - (p_1-p_3-q_1)^-} 
\nonumber \\
&& =  { 1 \over 2}{ \lambda^2 \over 4 \pi m^2}
{p_1^+ p_3^+ \over p_1^+ + p_3^+} {\theta(p_1^+ - p_3^+) \over 
p_1^+ - p_3^+}  \int_0^{p_1^+ - p_3^+} dq_1^+  
\Big [ {1 \over q_1^+ - p_1^+}  - { 1 \over q_1^+ + p_3^+} \Big ].
\label{lff1}
\end{eqnarray}

2) $p_1^+ < p_3^+$. 

The scattering amplitude (Fig. 1b) is  
\begin{eqnarray}
M_{fi} && = {1 \over 2}{ \lambda^2 \over 4 \pi } ~\theta(p_3^+ - p_1^+)~
\int_0^{p_3^+ - p_1^+} 
dq_1^+ 
 ~{ 1 \over q_1^+} ~{ 1 \over p_3^+ - p_1^+ - q_1^+}
\nonumber \\
&&~~~~~~~~~{ 1 \over p_3^- + p_2^- - p_1^- -p_2^- - q_1^- - (p_3-p_1-q_1)^-} 
\nonumber \\
&& =  { 1 \over 2}{ \lambda^2 \over 4 \pi m^2}
{p_1^+ p_3^+ \over p_1^+ + p_3^+} {\theta(p_3^+ - p_1^+) \over 
p_3^+ - p_1^+}  \int_0^{p_3^+ - p_1^+} dq_1^+  
\Big [ {1 \over q_1^+ - p_3^+}  - { 1 \over q_1^+ + p_1^+} \Big ].
\label{lff2}
\end{eqnarray}
We have used overall energy conservation $ p_1^- + p_2^- = p_3^- + p_4^-$ and
hence $ p_2^- - p_4^- = p_3^- - p_1^-$. 

We are interested in the forward scattering amplitude, i.e., in  $ \mid p_1^+
- p_3^+ \mid \rightarrow 0 $ limit. In this limit $q_1^+$ is very small 
compared to both $p_1^+$ and $p_3^+$ and it is legitimate to expand the integrands. 
We get,
\begin{eqnarray}
{ 1 \over q_1^+ - p_1^+ } - { 1 \over q_1^+ + p_3^+} && \approx - {p_1^+ +
p_3^+ \over p_1^+ p_3^+}, \nonumber \\
{ 1 \over q_1^+ - p_3^+ } - { 1 \over q_1^+ + p_1^+} && \approx - {p_1^+ +
p_3^+ \over p_1^+ p_3^+}.
\end{eqnarray}
Thus, in the forward scattering limit, we get,
\begin{eqnarray}
M_{fi} =- { 1 \over 2} {\lambda^2 \over 4 \pi m^2}.
\label{fw}
\end{eqnarray}
Alternatively, we can write the scattering amplitude as 
\begin{eqnarray}
M_{fi}= \frac{1}{2}\frac{\lambda^2}{4\pi}\int\limits_{0}^{1}dy\frac{1}
{y(1-y)t-m^2 + i \epsilon}
\end{eqnarray}
and calculate it explicitly:
\begin{eqnarray}
M_{fi}(t)=-{ 1 \over 2} {\lambda^2 \over 4 \pi}
\frac{1}{t\sqrt{{1 \over 4}-{m^2 \over t}}}\log \left(
\frac{2\sqrt{{1 \over 4}-{m^2 \over t}}-1}{2\sqrt{{1 \over 4} - {m^2 \over t}} 
+1}\right). \label{full}
\end{eqnarray}
In the forward scattering limit, one again finds the result (\ref{fw}). 
%%%%%%%%%%%%%%%%%%%%%%%%%%%%%%%%%%%%%%%%%%%%%%%%%%%%%%%%%%%%
\subsubsection{$s$-channel scattering}
%%%%%%%%%%%%%%%%%%%%%%%%%%%%%%%%%%%%%%%%%%%%%%%%%%%%%%%%%%%% 
For the $s$-channel scattering we have
\begin{eqnarray}
T_{fi} &&= { \lambda^2 \over 8 \pi} \int_0^{p_1^++p_2^+} ~ dq^+
{ 1 \over q^+ (p_1^+ + p_2^+ - q^+)} { 1 \over p_1^- + p_2^- - {m^2 \over
q^+} - {m^2 \over p_1^+ + p_2^+ - q^+}+ i \epsilon} \nonumber \\
&& = { \lambda^2 \over 8 \pi} \int_0^{p_1^++p_2^+} ~ dq^+
{ 1 \over q^+(p_1^+ + p_2^+)(p_1^- +p_2^-)- (q^+)^2(p_1^- + p_2^-)- m^2(p_1^+ + p_2^+) + i
\epsilon} \nonumber \\
&& = { \lambda^2 \over 8 \pi} \int_0^1 ~ dy ~ { 1 \over y(1-y)s - m^2 + i
\epsilon}. 
\end{eqnarray}
We have introduced $ s = (p_1^+ + p_2^+ )(p_1^- + p_2^-)$, $y={q^+ \over p_1^+
+ p_2^+}$.
An explicit evaluation leads to 
\begin{eqnarray}
{\rm Re} ~T_{fi} = - { \lambda^2 \over 4 \pi} ~ 
{ 1 \over s \sqrt{1 - { 4 m^2 \over
s}}}~ {\rm ln} { 1 - y_+ \over y_+}
\end{eqnarray}
where $ y_+ = { 1 \over 2} \left [ 1 + \sqrt{1 - 4 (m^2/ s)} \right ] $. 
 
%%%%%%%%%%%%%%%%%%%%%%%%%%%%%%%%%%%%%%%%%%%%%%%%%%%%%%%%%%
\subsection{Discretized 
formulation} 
%%%%%%%%%%%%%%%%%%%%%%%%%%%%%%%%%%%%%%%%%%%%%%%%%%%%%%%%%%%
\subsubsection{Periodic Boundary Condition}
%%%%%%%%%%%%%%%%%%%%%%%%%%%%%%%%%%%%%%%%%%%%%%%%%%%%%%%%%%%
In order to calculate the one-loop scattering amplitude in DLCQ perturbation 
theory for the $\lambda/(4!)^{-1}\phi^4$ (1+1) model with periodic boundary conditions, 
we need to derive the light front Hamiltonian 
with $O(\lambda^2)$ ZM effective interactions. However, since it was already
shown\cite{Harindranath:2000vf} that contributions from ZM induced 
effective interactions decouple in the continuum limit, we shall ignore these
contributions from the very beginning.   
The mode expansion for the normal mode field $\phi_n(x^-)$ is
\begin{eqnarray}
\phi_n(x^-) = {1 \over {\sqrt{2L}}}\sum_{k_n^+ > 0}{1 \over 
{\sqrt{k^+_n}}}
\left[a_n e^{-ikx} + a^{\dagger}_n
e^{ikx} \right].
\label{phiexp}
\end{eqnarray}
Here we have used the notation $kx \equiv {1\over 2}k_n^+x^-$ and   
$k_n^+={2\pi \over L}n, n=1,2,\dots \infty$. 

The scattering amplitude can be calculated by the old fashioned perturbation 
theory formula
\begin{eqnarray}
T_{fi} = \sum_{j}{{\langle {p}^\prime \vert H_{I}\vert j \rangle
\langle j \vert H_{I} \vert {p} \rangle}\over{p^- - p^-_{j}}},
\label{PTformula}
\end{eqnarray}
where $H_I$ denotes the interacting Hamiltonian. 
Using the formula (\ref{PTformula}) with $\vert p \rangle  
\rightarrow \vert p_1^+,p_2^+ \rangle $, $\vert 
p^\prime \rangle \rightarrow \vert p_3^+,p_4^+ \rangle $  
and with four-particle intermediate states, one finds the following
expression for the second-order 
normal mode scattering amplitude  
\begin{eqnarray}
T_{fi} = {{\delta_{p_4^++p_3^+,p_2^+ +p_1^+}\theta(p^+_3 - p^+_1)}
\over{(2L)^2\sqrt{p^+_4  p^+_3  
p^+_2  p^+_1 }}}{\lambda^2 \over 4}\sum_
{q_1^+}{1 \over{q_1^+(p^+_3-p^+_1-q_1^+)}}{1 \over{p^-_3 - p^-_1 - q_1^-
- (p_3- p_1-q_1)^-} }
%+(1 \leftrightarrow 3)
%\qquad \qquad \qquad \qquad \qquad 
\label{DLCQ4}
\end{eqnarray}
plus another term with $1 \leftrightarrow 3$. 
The above equation must be treated with care. Due to the presence of the
$\theta$-function, $p_1^+$ may approach $p_3^+$ to an arbitrary precision
but not to the exact value. 
In DLCQ, we have,
\begin{eqnarray}
t = (p_1^+ - p_3^+)(p_1^- - p_3^-) = - m^2 {(p_1^+ - p_3^+)^2 \over p_1^+
p_3^+} = -m^2 {(n_1 - n_3)^2 \over n_1 n_3},
\end{eqnarray} 
independent of $L$. For convenience, we set $m^2=1.0$ and without loss of
generality take $p_1^+ > p_3^+$.
The scattering amplitude (we have taken out the irrelevant factor
${\lambda^2 \over 8 \pi}$) is  
\begin{eqnarray}
M(t)={n_1 n_3 \over n_1 + n_3} { 1 \over n_1 - n_3} \sum_{n=1}^{n_1-n_3}
\Big [ { 1 \over n-n_1} - { 1 \over n+n_3} \Big ]. \label{sum}
\end{eqnarray}  
%%%%%%%%%%%%%%%%%%%%%%%%%%%%%%%%%%%%%%%%%%%%%%%%%%%%%%%%%%%%%%
\subsubsection{Anti Periodic Boundary Condition}
%%%%%%%%%%%%%%%%%%%%%%%%%%%%%%%%%%%%%%%%%%%%%%%%%%%%%%%%%%%%%%
With anti periodic boundary condition, the mode expansion for the field is
\begin{eqnarray}
\phi(x^-) = { 1 \over \sqrt{2 \pi}} \sum_{1,2,...} { 1 \over \sqrt{n}} \left
[ a_m e^{- { i \over 2} { \pi \over L} mx^-} +  a_m^\dagger 
e^{- { i \over 2} { \pi \over L} mx^-} \right ]. \label{phiap}
\end{eqnarray}
The scattering amplitude (we have taken out the irrelevant factor
${\lambda^2 \over 8 \pi}$) in the $t$-channel is  
\begin{eqnarray}
M(t)=2 {n_1 n_3 \over n_1 + n_3} { 1 \over n_1 - n_3} \sum_{n=1}^{n_1-n_3-1}
\Big [ { 1 \over n-n_1} - { 1 \over n+n_3} \Big ]. \label{sumap}
\end{eqnarray}

In the discretized version, the $s$-channel scattering amplitude is given by
\begin{eqnarray}
M(s) = 2 \sum_{n=1}^{n_{max}} { 1 \over
(2n-1) ({ 1 \over 2n_1-1})+{1 \over 2n_2 -1}) [ 2 n_{max} - (2n-1)] 
- 2 n_{max} + i \epsilon} \label{scsa}
\end{eqnarray}
where $ 2 n_{max} = (2n_1-1) +(2n_2-1)$ and $s = \Big [ (2n_1-1)+(2n_2-1)
\Big ] \Big [ { 1 \over 2n_1 -1}+{ 1 \over 2n_2-1} \Big ]$.
  
%%%%%%%%%%%%%%%%%%%%%%%%%%%%%%%
\subsection{Numerical Results}
%%%%%%%%%%%%%%%%%%%%%%%%%%%%%%%
%%%%%%%%%%%%%%%%%%%%%%%%%%%%%%%%%%%%%%%%%%%%%%%%%%
\subsubsection{Periodic Boundary Condition}
%%%%%%%%%%%%%%%%%%%%%%%%%%%%%%%%%%%%%%%%%%%%%%%%%%
Let us evaluate the 
scattering amplitude given in Eq. (\ref{sum}) in DLCQ. Note that the minimum
allowed value for $n_1$, $n_3$ is 1. Thus we start from $n_1=2$. In this case
$n_3=1$ and DLCQ gives the answer -1 for the scattering amplitude for
$t=-1/2$ which is obviously wrong. It is easy to check that 
for each $n_1$, since the maximum $n_3$ is $n_1-1$, the corresponding 
minimum $t$ is - ${ 1 \over n_1 (n_1-1)}$ 
and for this particular
$t$ DLCQ always gives the answer $-1$ for the scattering amplitude which is
wrong for finite $n_1$ but is correct for $ n_1 \rightarrow \infty$. The next
maximum value of $n_3$ is $n_1-2$ and we denote the corresponding $t$ by 
${\tilde t} = - { 4 \over n_1(n_1-2)}$. In table I we present the behavior
of $M({\tilde t})$ with $n_1$ as ${\tilde t} \rightarrow 0$. It is clear from 
Table I that DLCQ produces the correct answer which is $-1$ in our 
units, for the limit of forward scattering. Again, the limit may be
approached to an arbitrary numerical precision.

For a given $n_1$, we increase $n_3$  by steps of 2 and
study the behavior of $M(t)$ as a function of $t$ for small values of $t$. 
The result is plotted in Fig. 2. Recall that for $n_1=2$, $n_3=1$, $t=-1/2$ and
$M(t)=-1$. For $n_1=4$, $n_3=2$, $t=-1/2$ and $M(t)=-0.94$ which is 
close to the continuum limit ($-0.92$). Thus, for very small $n_1$, with periodic
boundary condition, the convergence is from below.
We can see that results for very 
small $n_1$ are affected by discretization but reliable results emerge
already for $n_1$=10. 
This is further confirmed by Fig. 3 where we present
the results for $n_1$=10, 20 and 30 and also present the continuum result
given in Eq. (\ref{full}) for comparison. In Fig. 4 we present the  
result for $n_1=2000$ and the continuum result. It is evident that 
DLCQ reproduces the continuum answer for the entire range of $t$ including 
the forward scattering limit $t=0$.   
%%%%%%%%%%%%%%%%%%%%%%%%%%%%%%%%%%%%%%%%%%%%%%%%%%%%%%%%%%%%
\subsubsection{Anti Periodic Boundary Condition}
%%%%%%%%%%%%%%%%%%%%%%%%%%%%%%%%%%%%%%%%%%%%%%%%%%%%%%%%%%%%
We evaluate the scattering amplitude given in Eq. (\ref{sumap}) for anti
periodic boundary condition in DLCQ. For the minimum value of $n_1=3$,
$n_3=1$, $ t=-4/3$, $M(t)=-3/4$ which is away from the continuum limit. For
$n_1=9$, $n_3=3$, $t=-4/3$,  $M(t)=-0.81$ which is closer to the continuum
limit ($-0.82$). Thus for very small $n_1$, with anti periodic boundary condition, the
convergence is from above. We can see that results for very 
small $n_1$ are affected by discretization but reliable results
emerge already for $n_1$=9. The behavior of $M(t)$ as a function of $t$ for
small values of $n_1$ is plotted in Fig. 5. In Fig. 6 we present the  
result for $n_1=2001$ and the continuum result. It is evident that  
DLCQ reproduces the continuum answer for the entire range of $t$ including
the forward scattering limit $t=0$ also for anti periodic boundary condition.
 
%%%%%%%%%%%%%%%%%%%%%%%%%%%%%%%%%%%%%%%%%%%%%%%%%%%%%%%%%%%%%%
\subsubsection{$s$-channel scattering}
%%%%%%%%%%%%%%%%%%%%%%%%%%%%%%%%%%%%%%%%%%%%%%%%%%%%%%%%%%%%%%
We choose antiperiodic boundary condition and evaluate the real 
part of the $s$-channel scattering amplitude 
given in Eq. (\ref{scsa}). For
$s=10$ and $4.2$, we start from small $n_1$ and solve for $n_2$ and
calculate the real part of the scattering amplitude. 
The real part of the amplitude converges rapidly in DLCQ to the continuum result with
increasing $n_1$, $n_2$ at fixed $s$ (which represents the continuum limit for
scattering problems in DLCQ). 
The results for the real part of the amplitude are presented in Tables II and III where the approach to continuum
limit is shown to be quicker for values of $s$ away from the threshold value
(4.0).

%%%%%%%%%%%%%%%%%%%%%%%%%%%%%%%%%%%%%%%%%%%%%%%%%%%%%%%%%%%
\section{Discussion, Summary and Conclusions}
%%%%%%%%%%%%%%%%%%%%%%%%%%%%%%%%%%%%%%%%%%%%%%%%%%%%%%%%%%%

The question whether DLCQ can produce the correct continuum limit
is nontrivial in 3+1 dimensions due to divergences, renormalization 
etc.. Two dimensional scalar field theory allows us to unambiguously answer 
this question.

It is worthwhile to contrast the calculations of mass spectra and scattering
amplitudes in DLCQ. For the bound state spectra one is solving for the
invariant mass for various values of $K$ and $K \rightarrow \infty $ gives
the continuum limit. Calculations of scattering amplitudes present a
different situation. For $s$-channel scattering we fixed $s$, picked an
$n_1$ and solved for $n_2$ and calculated the amplitude for these values of
the external discretized momenta. Then we increased $n_1$ and solved for
$n_2$ for the same value of $s$. By going to larger values of $n_1$ we
showed how the continuum limit is approached in DLCQ. For $t$-channel scattering
we fixed $n_1$, and for allowed values of $n_3$ such that $p_1^+ > p_3^+$ we
calculated the scattering amplitude as a function of $t$. For increasing
values of $n_1$ we showed how continuum limit was reached.    

We have provided details of the straightforward calculations in the continuum 
and DLCQ versions of light front perturbation theory for the one loop 
scattering diagram in two dimensional scalar field theory. We have 
shown that the continuum limit of DLCQ produces the
correct covariant limit for processes with $p^+=0$ exchange in the
$t$-channel. It is also important to demonstrate that
DLCQ can produce the absorptive part of the scattering amplitude
above the particle production threshold, a subject for future research.

%%%%%%%%%%%%%%%%%%%%%%%%%%%%%%%%
\acknowledgements
This work was supported in part by the U.S. Department 
of Energy, Grant No. DE-FG02-87ER40371, Division of High Energy and 
Nuclear Physics, by the VEGA Grant No. 
2/7119/2000 and by the International Institute of 
Theoretical and Applied Physics, Iowa State University, Ames, Iowa, U.S.A.  
%%%%%%%%%%%%%%%%%%%%%%%%%%%%%%%%%%%%%%%%%%%%%%%%%%%%%%%%%%%%

\vskip 1in
%\centering
\begin{tabular}{||c|c|c||}
\hline \hline
& &  \\
  $n_1$ &  $-{\tilde t}$ & $M({\tilde t})$\\
& & \\
\hline
   6 &   .166667 $\times 10^{0}$ &   -.980000 \\
   8 &   .833333 $ \times 10^{-1}$ &   -.989796 \\
  10 &   .500000 $ \times 10^{-1}$ &   -.993827 \\
  20 &   .111111 $ \times 10^{-1}$ &   -.998615 \\
  30 &   .476190 $ \times 10^{-2}$ &   -.999405 \\
 100 &   .408163 $ \times 10^{-3}$ &   -.999949 \\
 500 &   .160643 $ \times 10^{-4}$  &   -.999998 \\ 
1000 &   .400802 $ \times 10^{-5}$ &   -.999999 \\
\hline
\hline
\end{tabular}
%\end{table}
\vskip .5in
Table I. $M({\tilde t})$ versus ${\tilde t}$ in DLCQ. 
 For the definition of ${\tilde t}$, see the text. 
The correct answer is $-1$ in our choice of units.
\vskip 1in
%\centering
\begin{tabular}{||c|c|c|c||}
\hline \hline
& &  & \\
  $n_1$ &  $n_2$ & $s$  & Re ${ 1 \over 8 \pi} M(s)$\\
& &  & \\
\hline
   4 &  28  & 9.984 & .0207590 \\
   12 & 91 &  9.996 & .0210703 \\
   20 & 154 & 9.999 & .0211225 \\
   44 & 343 & 10.001 & .0211631 \\
   83 & 650 & 10.000 & .0211807 \\
  162 & 1272& 10.000 & .0211891 \\
  327 & 2571 & 10.000 & .0211941 \\ 
 ~ 650~ & ~5114~ & ~10.000~ & .0211962~ \\
\hline
\hline
\end{tabular}
%\end{table}
\vskip .5in
Table II. Real part of ${ 1 \over 8 \pi} M(s)$  in DLCQ. $ \epsilon = .0001$. 
% Absorptive part of ${ 1 \over 8 \pi} M(s)$ is -1591.55 
The continuum answer is  .0211985.
\vskip 1in
\begin{tabular}{||c|c|c|c||}
\hline \hline
& &  & \\
  $n_1$ &  $n_2$ & $s$  & Re ${ 1 \over 8 \pi} M(s)$\\
& &  & \\
\hline
   4 &  6  & 4.208 & -.00426 \\
   13 & 20 &  4.201 & .02612 \\
   29 & 45 &  4.202 & .03310 \\
   58 & 90 &  4.199 & .03580 \\
   115 & 179 &4.200 & .03716 \\
   230 & 358 & 4.200 & .03784 \\
   461 & 718 & 4.200 & .03818 \\ 
   923 & 1438 & 4.200 & .03834 \\
  ~1846~ &~ 2876~ & ~4.200~ & ~.03843~ \\
\hline
\hline
\end{tabular}
%\end{table}
\vskip .5in
Table III. Real part of ${ 1 \over 8 \pi} M(s)$  in DLCQ. $ \epsilon = .0001$. 
 
The continuum answer is  .03851.
\eject
\begin{center}
{\bf Figures} 
\end{center}
\noindent 
\vskip .25in
\begin{center}
%
\vskip .2in
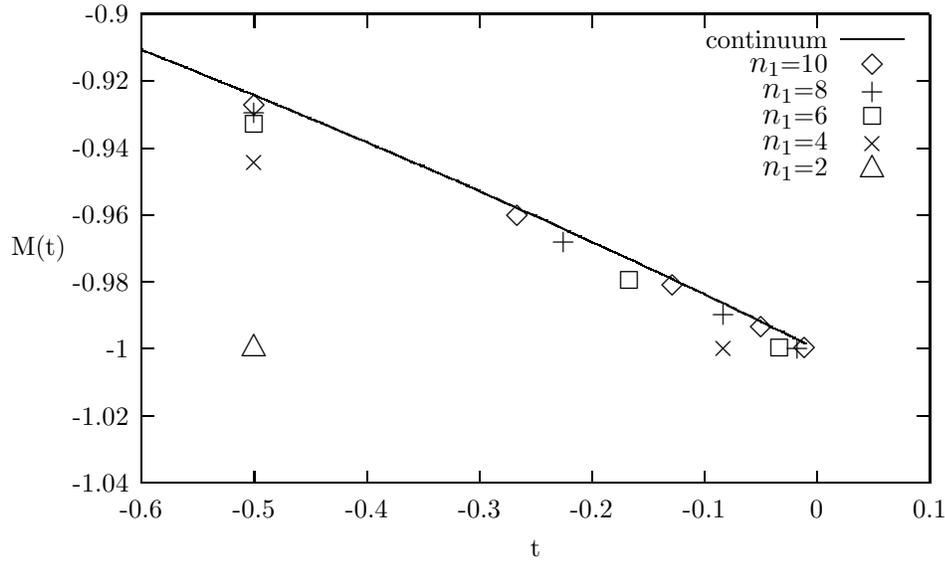
Fig. 2. $M(t)$ versus $t$ in DLCQ for $n_1$=2, 4, 6, 8, and 10 plotted for
small $t$.
\vskip 1in
% GNUPLOT: LaTeX picture
\setlength{\unitlength}{0.240900pt}
\ifx\plotpoint\undefined\newsavebox{\plotpoint}\fi
\sbox{\plotpoint}{\rule[-0.200pt]{0.400pt}{0.400pt}}%
%
\vskip .2in
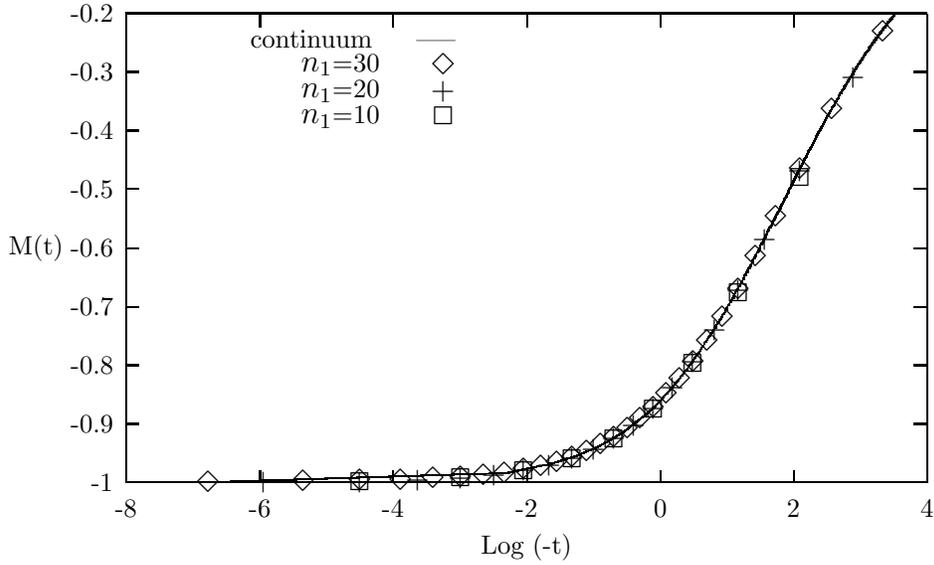
Fig. 3. $M(t)$ versus $Log (-t)$ in DLCQ for $n_1$=10, 20 and 30 compared with the
continuum result.
\vskip 1in
% GNUPLOT: LaTeX picture
\setlength{\unitlength}{0.240900pt}
\ifx\plotpoint\undefined\newsavebox{\plotpoint}\fi
\sbox{\plotpoint}{\rule[-0.200pt]{0.400pt}{0.400pt}}%
%
\vskip .2in
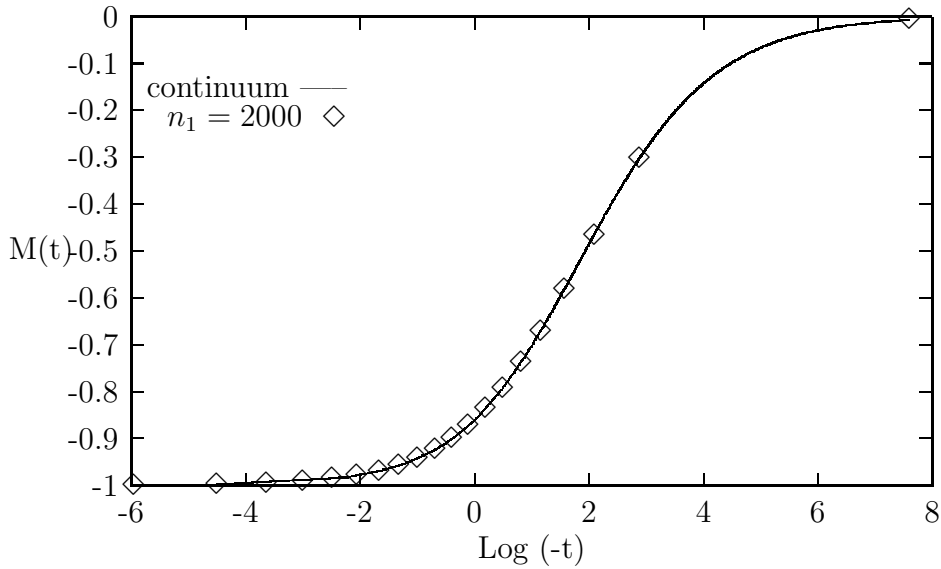
Fig. 4. $M(t)$ versus $Log (-t)$ for $n_1=2000$ compared with the continuum result.
\vskip 1in
% GNUPLOT: LaTeX picture
\setlength{\unitlength}{0.240900pt}
\ifx\plotpoint\undefined\newsavebox{\plotpoint}\fi
\sbox{\plotpoint}{\rule[-0.200pt]{0.400pt}{0.400pt}}%
%
\vskip .2in
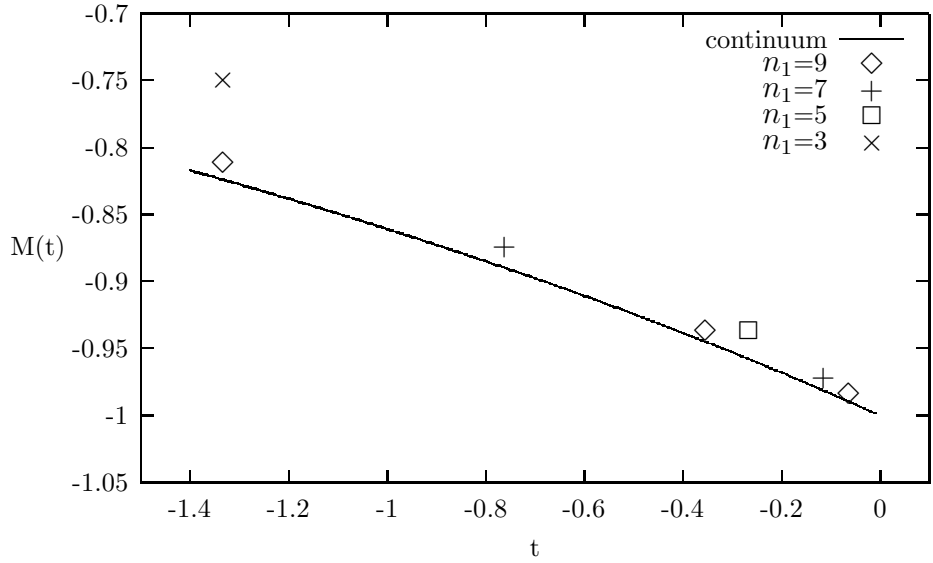
Fig. 5. $M(t)$ versus $t$ in DLCQ for $n_1$=3, 5, 7, and 9 plotted for
small $t$ (anti-periodic boundary condition). 
\vskip 1in
% GNUPLOT: LaTeX picture
\setlength{\unitlength}{0.240900pt}
\ifx\plotpoint\undefined\newsavebox{\plotpoint}\fi
\sbox{\plotpoint}{\rule[-0.200pt]{0.400pt}{0.400pt}}%
%
\vskip .2in
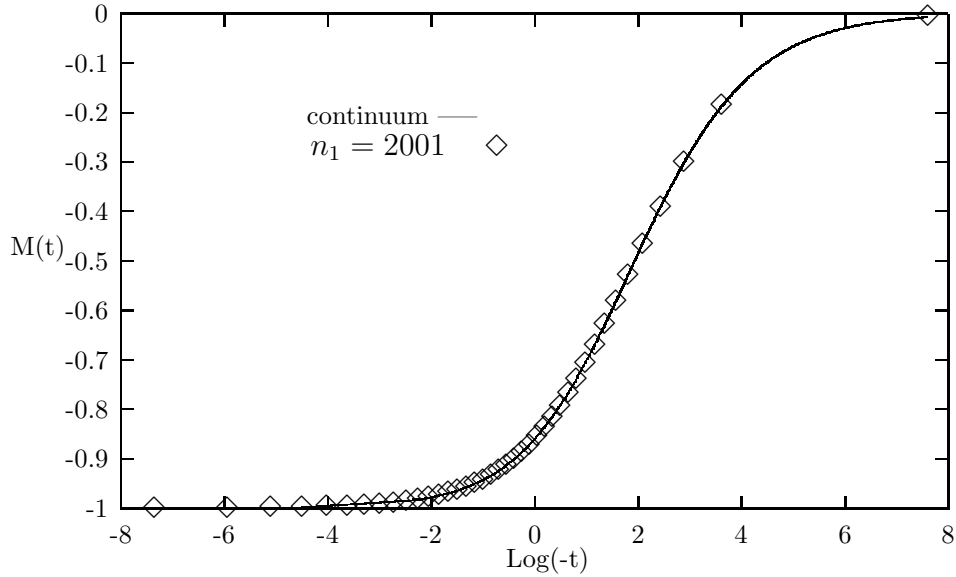
Fig. 6.  $M(t)$ versus $Log (-t)$ for $n_1=2001$ compared with the continuum 
result (anti-periodic boundary condition).

\begin{thebibliography}{99}


%\cite{Bjorken:1971ah}
\bibitem{Bjorken:1971ah}
J.~D.~Bjorken, J.~B.~Kogut and D.~E.~Soper,
%`Quantum Electrodynamics At Infinite Momentum: Scattering From An External
%Field,''
Phys.\ Rev.\ D {\bf 3}, 1382 (1971).
%%CITATION = PHRVA,D3,1382;%%

%\cite{Chang:1973qi}
\bibitem{Chang:1973qi}
S.~Chang and T.~Yan,
%`Quantum Field Theories In The Infinite Momentum Frame. 2. Scattering
%Matrices Of Scalar And Dirac Fields,''
Phys.\ Rev.\ D {\bf 7}, 1147 (1973).
%%CITATION = PHRVA,D7,1147;%%

%\cite{Suzuki:1976xb}
\bibitem{Suzuki:1976xb}
T.~Suzuki, S.~Tameike and E.~Yamada,
%`Some Undesirable Features Of Quantum Field Theory On A Null Plane,''
Prog.\ Theor.\ Phys.\  {\bf 55}, 922 (1976).
%%CITATION = PTPKA,55,922;%%

%\cite{Ji:1992xr}
\bibitem{Ji:1992xr}
C.~R.~Ji and Y.~Surya,
%`Calculation of scattering with the light cone two body equation in phi**3
%theories,''
Phys.\ Rev.\ D {\bf 46}, 3565 (1992).
%%CITATION = PHRVA,D46,3565;%%

%\cite{Fuda:1991nn}
\bibitem{Fuda:1991nn}
M.~G.~Fuda,
%`Angular momentum and light front scattering theory,''
Phys.\ Rev.\ D {\bf 44}, 1880 (1991).
%%CITATION = PHRVA,D44,1880;%%


%\cite{Maskawa:1976ky}
\bibitem{Maskawa:1976ky}
T.~Maskawa and K.~Yamawaki,
%`The Problem Of P+ = O Mode In The Null Plane Field Theory And Dirac's
%Method Of Quantization,''
Prog.\ Theor.\ Phys.\  {\bf 56}, 270 (1976).
%%CITATION = PTPKA,56,270;%%


%\cite{Casher:1976ae}
\bibitem{Casher:1976ae}
A.~Casher,
%`Gauge Fields On The Null Plane,''
Phys.\ Rev.\ D {\bf 14}, 452 (1976).
%%CITATION = PHRVA,D14,452;%%


%\cite{Thorn:1978kx}
\bibitem{Thorn:1978kx}
C.~B.~Thorn,
%`On The Derivation Of Dual Models From Field Theory.  2,''
Phys.\ Rev.\ D {\bf 17}, 1073 (1978).
%%CITATION = PHRVA,D17,1073;%%


%\cite{Pauli:1985pv}
\bibitem{Pauli:1985pv}
H.~C.~Pauli and S.~J.~Brodsky,
%`Solving Field Theory In One Space One Time Dimension,''
Phys.\ Rev.\ D {\bf 32}, 1993 (1985).
%%CITATION = PHRVA,D32,1993;%%

%\cite{Pauli:1985ps}
\bibitem{Pauli:1985ps}
H.~C.~Pauli and S.~J.~Brodsky,
%`Discretized Light Cone Quantization: Solution To A Field Theory In One
%Space One Time Dimensions,''
Phys.\ Rev.\ D {\bf 32}, 2001 (1985).
%%CITATION = PHRVA,D32,2001;%%


%\cite{Brodsky:1998de}
\bibitem{Brodsky:1998de}
S.~J.~Brodsky, H.~Pauli and S.~S.~Pinsky,
%Quantum chromodynamics and other field theories on the light cone,''
Phys.\ Rept.\ {\bf 301}, 299 (1998)
[hep-ph/9705477]. 
%%CITATION = HEP-PH 9705477;%%
For most recent work in Discrete Light Cone Quantization
see J. Hiller, {\it Application of Discrete Light Cone Quantization to
Yukawa Theory in Four Dimensions},
[hep-ph/0010061] and references therein.



%\cite{Hiller:2000vi}
\bibitem{Hiller:2000vi}
J.~R.~Hiller,
%`Nonperturbative calculation of scattering amplitudes,''
hep-ph/0007231.
%%CITATION = HEP-PH 0007231;%%



%\cite{Chakrabarti:2000cg}
\bibitem{Chakrabarti:2000cg}
D.~Chakrabarti, A.~Mukherjee, R.~Kundu and A.~Harindranath,
%`A numerical experiment in DLCQ: Microcausality, continuum limit and all
%that,''
Phys.\ Lett.\ B {\bf 480}, 409 (2000)
[hep-th/9910108].
%%CITATION = HEP-TH 9910108;%%



%\cite{Harindranath:2000vf}
\bibitem{Harindranath:2000vf}
A.~Harindranath, L.~Martinovic and J.~P.~Vary,
%`Compactification near and on the light front,''
Phys.\ Rev.\ D {\bf 62}, 105015 (2000)
[hep-th/9912085].
%%CITATION = HEP-TH 9912085;%%

%\cite{Hellerman:1999yu}
\bibitem{Hellerman:1999yu}
S.~Hellerman and J.~Polchinski,
%`Compactification in the light like limit,''
Phys.\ Rev.\ D {\bf 59}, 125002 (1999)
[hep-th/9711037].
%%CITATION = HEP-TH 9711037;%%

%\cite{Taniguchi:2001cb}
\bibitem{Taniguchi:2001cb}
M.~Taniguchi, S.~Uehara, S.~Yamada and K.~Yamawaki,
%`Does DLCQ S-matrix have a covariant continuum limit?,''
Mod. Phys. Lett. A {\bf 16}, 2177 (2001)
[hep-th/0106167].
%%CITATION = HEP-TH 0106167;%%


\end{thebibliography}
\end{document}